\documentstyle[prb,aps,epsfig,draft,floats,amsfonts]{revtex}

\begin{document}

\twocolumn \psfull \draft

\wideabs{

\title{Statistical Properties of Eigenstates in
three-dimensional Mesoscopic Systems with Off-diagonal or Diagonal
Disorder}
\author{Branislav K. Nikoli\' c$^\ddag$}
\address{Department of Physics and Astronomy,
SUNY at Stony Brook, Stony Brook, New York 11794-3800}

\maketitle

\begin{abstract}
The statistics of eigenfunction amplitudes are studied in
mesoscopic disordered electron systems of finite size. The exact
eigenspectrum and eigenstates are obtained by solving numerically
Anderson Hamiltonian on a three-dimensional lattice for different
strengths of disorder introduced either in the potential on-site
energy (``diagonal'') or in the hopping integral
(``off-diagonal''). The samples are characterized by the exact
zero-temperature conductance computed using real-space Green
function technique and related Landauer-type formula. The
comparison of eigenstate statistics in two models of disorder
shows sample-specific details which are not fully taken into
account by the conductance, shape of the sample and
dimensionality. The wave function amplitude distributions for the
states belonging to different transport regimes within the same
model are contrasted with each other as well as with universal
predictions of random matrix theory valid in the infinite
conductance limit.
\end{abstract}

}

\narrowtext

The disorder induced localization-delocalization (LD) transition
in solids has been one of the most vigorously pursued problems in
condensed matter physics since the seminal work of
Anderson.~\cite{anderson} In thermodynamic limit, strong enough
disorder generates a zero-temperature critical point in $d>2$
dimensions~\cite{kramer} as a result of quantum interference
effects (in $d \le 2$ even weakly disordered metal turns into an
Anderson insulator for sufficiently large sample size). Thus,
research in the ``pre-mesoscopic'' era~\cite{lee} was mostly
directed toward the viewpoint provided by the theory of critical
phenomena.~\cite{gang4} The advent of mesoscopic quantum
physics~\cite{mqp} has unearthed large fluctuations, induced by
quantum coherence and randomness of disorder,~\cite{jansen} of
various physical quantities~\cite{lerner} (e.g., conductance,
local density of states, current relaxation times, etc.), even
well into the delocalized phase. Thus, complete understanding of
the LD transition requires to examine full distribution functions
of relevant quantities.~\cite{shapiro} Especially interesting are
the deviations of their asymptotic tails, a putative signature of
incipient localization,~\cite{lerner} from the (usually) Gaussian
distributions expected in the limit of infinite dimensionless
conductance $g=G/G_Q$ ($G_Q=2e^2/h$ is the conductance quantum).

This paper presents the study of such type---numerical
computation of the statistics of eigenfunction amplitudes in
finite-size three-dimensional (3D) nanoscale (composed of
$\sim1000$ atoms)
 mesoscopic disordered conductors. The 3D conductors are often
``neglected'' in favor of the more popular and tractable
playgrounds---two-dimensional systems (2D), where one can study
states resembling 3D critical wave functions in a wide range of
systems sizes and disorder strengths,~\cite{mirlin,muller} or
quasi one-dimensional systems~\cite{rare_onedim} where analytical
techniques~\cite{efetovbook,dmpk} can handle even
non-perturbative phenomena~\cite{fmmodes} (like the ones at small
$g$). For example, in $d=2+\epsilon$ dimensions LD transition
occurs at weak disorder (weak-coupling regime of the corresponding
field-theoretical description~\cite{efetovbook}), while in $d \ge
3$ small parameter needed for analytical treatment is lacking. In
3D systems critical eigenfunctions, exhibiting multifractal
scaling,~\cite{jansen} are expected only at the mobility edge
$E_c$ which separates extended and localized states inside the
energy band.

The essential physics of disordered conductors is captured by
studying the quantum dynamics of a non-interacting
(quasi)particle in a random potential. This problem
is classically non-integrable, thereby exhibiting quantum chaos.
The concepts unifying disordered electron physics with the
standard `clean' (i.e., without stochastic disorder) examples of
quantum chaos~\cite{chaos} come from statistical approaches to
the properties of energy spectrum and corresponding eigenstates,
which cannot be computed analytically. While level statistics of
disordered systems have been explored to a great
extent,~\cite{mont,ghur} investigation of the statistics of
eigenfunctions has been initiated only recently.~\cite{mirlin}
These studies are not only revealing peculiar spectral properties
of random Hamiltonians, but are relevant for the thorough
understanding of various unusual features of quantum transport in
diffusive metallic conductors (including the ones which are
proximity coupled to a superconductor~\cite{ostrovsky}). The
standard examples are long-time tails in the relaxation of
current~\cite{muz} or log-normal tails (in $d=2+\epsilon$) of the
distribution function of mesoscopic conductances.~\cite{lerner}
Distribution of eigenfunction amplitudes is found to be relevant
for tunneling measurements on quantum dots probing the coupling
to external leads, which depends sensitively on the local
properties of wave functions.~\cite{dots}
\begin{figure}
\centerline{
\psfig{file=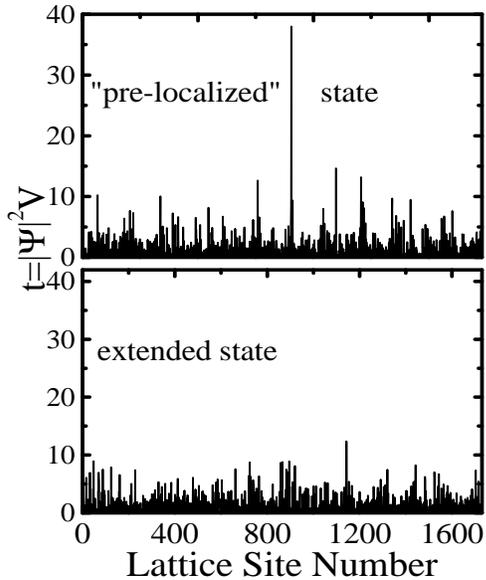,height=3.0in,width=2.5in,angle=0} }
\vspace{0.2in} \caption{An example of eigenstates in the band
center of a delocalized phase. The average conductance at half
filling is $g(E_F=0)\approx17$, entailing anomalous rarity of the
``pre-localized'' states. The disordered conductor is modeled by
an Anderson model with diagonal disorder $W_{\rm DD}=4$ on a simple cubic
lattice $12^3$. For plotting of the eigenfunction values in 3D,
the sites ${\bf m}$ are mapped onto the lattice site numbers $\in
\{1,...,1728\}$ in a lexicographic order, i.e., ${\bf
m}\equiv(m_x,m_y,m_z) \mapsto 144(m_x-1)+12(m_y-1)+m_z$.}
\label{fig:state}
\end{figure}
Experiments which are the closest to directly delving into the
microscopic structure of quantum chaotic or disordered wave
functions exploit the correspondence between the Schr\" odinger
and Maxwell equations in microwave cavities.~\cite{sridhar}

The study of fluctuations and correlations of eigenfunction
amplitudes~\cite{mirlin,mirlinprb} in mesoscopic systems has led
to a concept of the so-called ``pre-localized''
states.~\cite{lerner,muz,efetov} In 3D delocalized phase, this
notion refers to the states which have sharp amplitude peaks on
the top of an extended background.~\cite{crit} These kind of
states appear even in the diffusive, $\ell \ll L < \xi$, metallic
($g \gg 1$) regime, but are anomalously rare in such samples (here
$\ell$ is the elastic mean free path and $\xi$ is the
localization length). In order to get ``experimental'' feeling for
the structure of states with unusually high amplitude spikes, an
example is given in Fig.~\ref{fig:state}; this state is found in
a special realization of quenched disorder (out of many randomly
generated impurity configurations) inside the sample
characterized by large average conductance. Thus, pre-localized
states are putative precursors of LD transition and determine
asymptotics of some of the distribution
functions~\cite{jansen,mirlin} studied in open or closed
mesoscopic systems. In $d \leq 2$, where all states are
considered to be localized,~\cite{gang4} pre-localized states
have anomalously short localization radius~\cite{mirlin} when
compared to ``ordinary'' localized states in low-dimensional
systems. They underlie~\cite{efetov} the multifractal scaling  in
weakly localized ($g \gg 1$) 2D conductors of size $L$ smaller
than exponentially large $\xi$ (which plays the role of a phase
transition correlation length~\cite{jansen} $\xi_c=\xi$ in $d \leq
2$). In 3D, the correlation length $\xi_c$ (defined as the size
of a hypercube for which~\cite{jansen} $g(\xi_c)={\mathcal O}(1)$)
is always microscopic ($\xi_c \sim \lambda_F$) in good metallic
samples, and no multifractal scaling is expected. The appearance
of small regions inside disordered solids where eigenstates can
have large amplitudes seems to be a ``strongly pronounced''
analog~\cite{muller,sridhar} of the phenomenon of
scarring~\cite{heller} (anomalous enhancement or suppression of
quantum chaotic wave function intensity on the unstable periodic
orbits in the corresponding classical system) introduced in the
guise of generic quantum chaos.

In general, the study of properties of wave functions on a scale
smaller than $\xi$ should probe quantum effects causing evolution
of extended into localized states upon approaching the LD
critical point. In the marginal two-dimensional case, the
divergent (in the limit $L \rightarrow \infty$) weak localization
(WL) correction~\cite{wl} to the semiclassical Boltzmann
conductivity provides an explanation of localization in terms of
the interference between two amplitudes to return to initial
point along the same classical path in the opposite
directions.~\cite{wl_timerev} This simple quantum interference
effect leads to a coherent backscattering (i.e., suppression of
conductivity) in a time-reversal invariant systems without
spin-orbit interaction. However, in 3D systems WL correction is
not ``strong'' enough to provide a full microscopic picture of
complicated quantum interference processes which are responsible
for LD transition, and facilitate the expansion of quantum
intuition.

\begin{figure}
\centerline{ \psfig{file=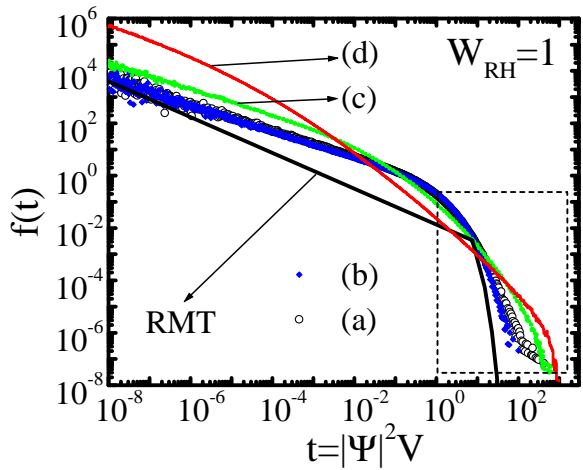,height=3.0in,angle=-90} }
\vspace{0.2in} \caption{Statistics of wave function intensities
in the RH Anderson model,  with $W_{\text{RH}}=1$, on a simple
cubic lattice  $16^3$. The distribution function
$f(t)$ in Eq.~(\ref{eq:ft}) is computed for the states around the
following energies: (a) $E=0$, (b) $E=1.5$, (c) $E=2.55$, and (d)
$E=2.75$. Disorder averaging is performed over
$N_{\text{Ens}}=40$ different samples. The Porter-Thomas
distribution~(\ref{eq:porter}) is denoted by RMT. The part of the
distributions inside the dashed box in enlarged on
Fig.~\ref{fig:ftrh2}.} \label{fig:ftrh1}
\end{figure}

\begin{figure}
\centerline{ \psfig{file=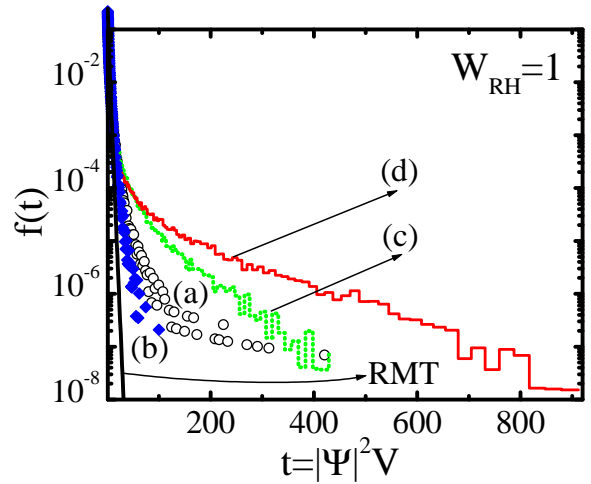,height=3.0in,angle=-90} }
\vspace{0.2in} \caption{Statistics of wave function intensities
in the RH Anderson model,  with $W_{\text{RH}}=1$, on a simple
cubic lattice with $N=16^3$ sites. This Figure plots the same
distributions $f(t)$ as the ones plotted in Fig.~\ref{fig:ftrh1},
in the range defined by the dashed square in Fig.~\ref{fig:ftrh1}.
The same labels apply to both Figures.} \label{fig:ftrh2}
\end{figure}

\begin{figure}
\centerline{
\psfig{file=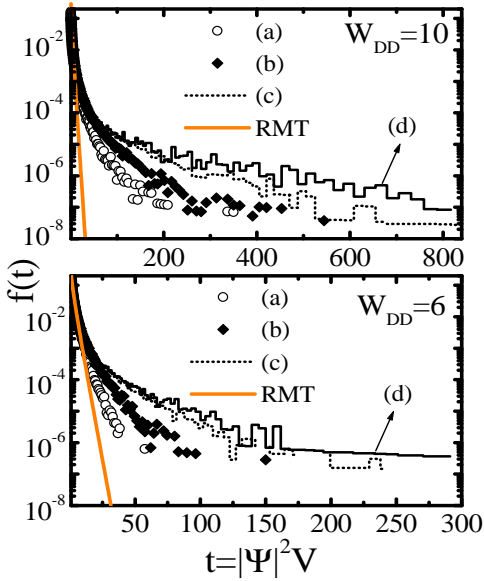,height=3.0in,width=2.5in,angle=0} }
\vspace{0.2in} \caption{Statistics of wave function intensities in
the DD Anderson model on a simple cubic lattice $16^3$. The
distribution function $f(t)$, Eq.~(\ref{eq:ft}), is computed for
the states around following energies. Upper panel,
$W_{\text{DD}}=10$:  (a) $E=0$, (b) $E=6.0$, (c) $E = 7.45$, and
(d) $E=7.85$. Lower panel, $W_{\text{DD}}=6$: (a) $E=0$, (b)
$E=4.1$, (c) $E = 6.56$, and (d) $E=6.7$. Disorder averaging is
performed over $N_{\text{Ens}}=40$ different samples. The
Porter-Thomas distribution~(\ref{eq:porter}) is denoted by RMT.}
\label{fig:ftdd}
\end{figure}

The paper presents the statistics of eigenfunction intensities
$|\Psi_\alpha({\bf r})|^2$ in isolated 3D mesoscopic conductors
characterized by two different types of microscopic disorder.
Numerical methods employed here make it possible to treat
phenomena in both semiclassical (described by Bloch-Boltzmann
formalism and perturbative quantum corrections~\cite{efetovbook})
and fully quantum transport regime (dominated by non-perturbative
effects, where semiclassical concepts, like $\ell$, loose their
meaning~\cite{allen}), as well as in the crossover realm. Since
mesoscopic physics has provided efficient techniques~\cite{datta}
for ``measuring'' exactly the transport properties of finite-size
samples on the computer, this study connects the eigenstates
statistics of a closed sample to its zero-temperature
conductance. The statistical properties of eigenstates are
described by the disorder-averaged distribution
function~\cite{fmmodes,efetov}
\begin{equation}\label{eq:ft}
    f(t)=\frac{1}{\rho(E) N} \left \langle \sum_{{\bf r},\alpha}
    \delta(t-|\Psi_\alpha({\bf r})|^2 V) \delta(E-E_\alpha)
  \right \rangle,
\end{equation}
on $N$ discrete points ${\bf r}$ inside a sample of volume $V$.
Here $\rho(E) = \langle \sum_\alpha \delta(E-E_\alpha) \rangle$ is
the mean level density at energy $E$. Averaging over disorder is
denoted by $\langle \ldots \rangle$. Normalization of eigenstates
gives $ \bar{t} = \int d t \, t \, f(t)=1$. A finite-size
disordered sample is modeled by a tight-binding Hamiltonian (TBH)
with nearest neighbor hopping integral $t_{{\bf m} {\bf n}}$
\begin{equation}\label{eq:tbh}
  \hat{H}=\sum_{\bf m} \varepsilon_{\bf m}  |{\bf m} \rangle \langle {\bf m}| +
  \sum_{\langle {\bf m},{\bf n} \rangle} t_{{\bf m} {\bf n}} |{\bf m} \rangle \langle {\bf
  n}|,
\end{equation}
on a simple cubic lattice $16 \times 16 \times 16$ of lattice
constant $a$. Each site ${\bf m}$ contains a single $s$-orbital
$\langle {\bf r} | {\bf m} \rangle = \psi({\bf r}-{\bf m})$.
Periodic boundary conditions are chosen in all directions. In a
random hopping (RH) model the disorder is introduced by taking the
off-diagonal matrix elements to be a uniformly distributed random
variable,~\cite{viktor} $1-2 W_{\text{RH}} < t_{{\bf m} {\bf n}}
< 1$, while diagonal elements are zero $\varepsilon_{\bf m}=0$.
The strength of the disorder is measured by $W_{\text{RH}}$. The
other system studied is described by a diagonally disordered (DD)
Anderson model with potential energy $\varepsilon_{\bf m}$ on
site ${\bf m}$ drawn from the uniform distribution,
$-W_{\text{DD}}/2 < \varepsilon_{\bf m} < W_{\text{DD}}/2$, and
$t_{{\bf m} {\bf n}}=1$ is the unit of energy. The
Hamiltonian~(\ref{eq:tbh}) is a real symmetric matrix because
time-reversal symmetry is assumed. The results for $f(t)$ in the
samples described by the RH  and DD Anderson models are shown on
Figs.~\ref{fig:ftrh1},~\ref{fig:ftrh2} and Fig.~\ref{fig:ftdd},
respectively. Although some of the samples are characterized by
similar values of conductance, the eigenstates in the two models
show different statistical behavior.
In what follows the meaning of these findings is explained in the
context of statistical approach to quantum systems with
non-integrable classical dynamics. In particular, the results are
compared to the universal predictions of random matrix theory
(RMT).

In the statistical approach~\cite{ghur} of RMT, the Hamiltonian of
a quantum chaotic system is replaced~\cite{foot} by a random
matrix drawn from an ensemble defined by the symmetries under
time-reversal and spin-rotation. This leads to the Wigner-Dyson
(WD) statistics for eigenvalues and Porter-Thomas (PT)
distribution for eigenfunction intensities. For the Gaussian
orthogonal ensemble (GOE), relevant for studies of
time-reversal-invariant Hamiltonians like~(\ref{eq:tbh}), the PT
distribution is given by
\begin{equation}\label{eq:porter}
  f_{\text{PT}}(t)=\frac{1}{\sqrt{2 \pi t}} \exp(-t/2).
\end{equation}
The function $f_{\text{PT}}(t)$ is plotted as a reference
in Figs.~\ref{fig:ftrh1},~\ref{fig:ftrh2} and Fig.~\ref{fig:ftdd}.
The predictions of RMT are universal, depending only on the
symmetry properties of the relevant ensemble. They apply to the
statistics of real disordered systems~\cite{gorkov} in the limit
$g \rightarrow \infty$ ($g=\pi E_{\text{Th}}/\Delta$, where
$\Delta=1/\rho(E)$ is the mean energy level spacing and
$E_{\text{Th}} = \hbar {\mathcal D}/L^2$ is the Thouless energy,
set by the classical diffusion across a sample of size $L$ with
diffusion constant ${\mathcal D}$). The spectral correlations in
RMT are determined by logarithmic level repulsion which is
independent of true dynamics.~\cite{ghur}
All sample-specific details are absorbed into the mean level
spacing~\cite{ghur} $\Delta$. Also, the level correlations are
independent of the eigenstate correlations. The RMT
answer~(\ref{eq:porter}) for the distribution
function~(\ref{eq:ft}) was derived by Porter and
Thomas~\cite{thomas} by assuming that the
coordinate-representation eigenstate $\langle {\bf r} |
\Psi_\alpha \rangle$ in a disordered (or classically chaotic
system) is a Gaussian random variable. The behavior of
$\Psi_\alpha({\bf r})$, even within the framework of RMT, is
simple only in the systems with unbroken or completely broken
time-reversal symmetry [the only difference between the two
limiting ensembles is the functional form of $f(t)$].~\cite{falko}
Thus, RMT implies statistical equivalence of eigenstates which
equally test the random potential all over the sample---typical
wave function has more or less uniform amplitude $1/\sqrt{V}$, up
to inevitable Gaussian fluctuations.

Microscopic theory brings corrections to the RMT results in the
case of samples with finite $g$. In the finite-size systems level
statistics follow RMT predictions in the ergodic regime, i.e., on
the energy separation scale smaller than $E_{\text{Th}}$.
Non-universal corrections to the spectral
statistics~\cite{andreev} or eigenfunction
statistics~\cite{mirlin,efetov,prigodin} (which describe the
long-range correlations of wave functions) depend on
dimensionality, shape of the sample, and conductance $g$. These
deviations from RMT predictions grow with increasing disorder
(i.e., lowering of $g$). At the LD transition wave functions
acquire multifractal properties, while the critical level
statistics become scale-independent.~\cite{shapiro_rmt} For
strong disorder or, at fixed disorder, for energies $|E|$ above
the mobility edge $|E_c|$, wave functions are exponentially
localized. The simple (and usually invoked) picture is that of a
wave function which decays as $\Psi(r)=p(r)\exp (-r/\xi)$ from
its maximum centered  at some point inside the sample of size $L >
\xi$. Here $p(r)$ is a random function and approximately
spherical symmetry of decay is assumed. Since two states close in
energy are localized at different points in space, there is
almost no overlap between them. Therefore, the levels become
uncorrelated and obey Poisson statistics. If $p(r)=c$ is
simplified  to a normalization constant, the distribution
function of intensities is given by
\begin{eqnarray} \label{eq:ftexp}
  f_{\xi}(t) & = & \frac{4\pi}{V} \int\limits_0^{L/2} dr\, r^2 \delta(t-|\Psi(r)|^2
  V)= \frac{\pi \xi^3}{4V} \frac{\ln^2(c^2 V/t)}{t}, \nonumber \\
  c^2 & = & \frac{2}{\pi \xi^3} \left [ 1-\left(1+\frac{L}{\xi}+\frac{L^2}{2\xi^2}\right)
  e^{-L/\xi} \right ]
  ^{-1},
\end{eqnarray}
where a spherically symmetric sample of radius $L/2$ is assumed.
In the localized phase $\xi \ll L$, $f_\xi(t)$ is expected to be
insensitive to the assumed shape of the sample, and is determined
by the ratio of these two relevant length scales.

The distribution function $f(t)$ is equivalently given in term of
its moments $b_q=\int d t \,t^q \, f(t)$ (this statement is not
rigorous since examples of different distribution functions which
possess exactly the same sets of moments are encountered in
various statistical problems~\cite{shapiroprl,bouchaud}). For GOE,
the PT distribution~(\ref{eq:porter}) has moments
$b_q^{\text{PT}} = 2^q V^{-q+1} \Gamma(q+1/2) / \Gamma(1/2)$.
They are related to the moments $I_{\alpha}(q)=\int d {\bf r} \,
|\Psi_\alpha({\bf r})|^{2q}$ of the wave function intensity.
\begin{figure}
\centerline{
\psfig{file=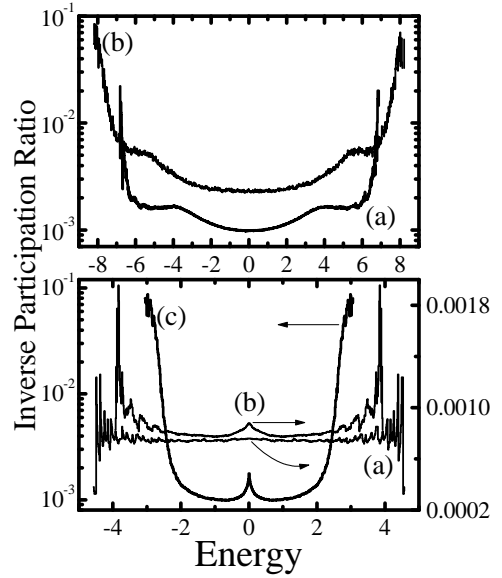,height=3.0in,width=2.5in,angle=0} }
\vspace{0.2in} \caption{Inverse Participation Ratio $I(2)$
,averaged over both 40 different conductors and small energy
bins,  of eigenstates in the RH and DD Anderson models on a simple
cubic lattice with $N=16^3$ sites. Top: diagonal disorder with
(a) $W_{\text{DD}}=6$, and (b) $W_{\text{DD}}=10$. Bottom:
off-diagonal disorder with (a) $W_{\text{RH}}=0.25$, (b)
$W_{\text{RH}}=0.375$, and (c) $W_{\text{RH}}=1$.} \label{fig:ipr}
\end{figure}
In the universal regime $g \rightarrow \infty$ wave functions
cover the whole volume with only short-range correlations (on the
scale $|{\bf r}_1-{\bf r}_2| \lesssim \ell$) persisting between
$\Psi_\alpha({\bf r}_1)$ and $\Psi_\alpha({\bf r}_2)$. This means
that integration in the definition of $I_{\alpha}(q)$ provides
self-averaging, and $I_{\alpha}(q)$ does not fluctuate in the
universal limit,~\cite{prigodin} i.e.,
$I_{\alpha}(q)=b_q^{\text{PT}}$. On the other hand, at finite $g$
spatial correlations of wave function amplitudes at distances
comparable to the system size are non-negligible. Therefore,
$I_{\alpha}(q)$ fluctuates from state to state and from sample to
sample.~\cite{mirlinprb,evers} Although these long-range spatial
correlations necessitate to study the full distribution
function~\cite{prigodin,evers} of $I_{\alpha}(q)$, for the
subsequent analysis in this study it is enough to use an ensemble
average of $I_{\alpha}(q)$, i.e., following Wegner~\cite{wegner}
\begin{equation} \label{eq:ipr}
I(q)= \Delta \left \langle \sum_{{\bf r},\alpha} \,
|\Psi_{\alpha}({\bf r})|^{2q} \delta(E-E_{\alpha}) \right \rangle.
\end{equation}
The moment $I_{\alpha}(2)$ is usually called inverse
participation ratio (IPR). It is a one-number measure of the
degree of localization (i.e., it measures the portion of space
where the amplitude of the wave function differs markedly from
zero). This becomes obvious from the scaling properties of the
average moments $I(q)$ with respect to the system size
\begin{equation} \label{eq:iprscaling}
I(q) \propto \cases{L^{-d(q-1)}  \ \ \ \ \ \ \ \ {\text{ metal}},
          \cr L^0  \ \ \ \ \ \ \ \ \ \ \ \ \ {\text{insulator}}, \cr
          L^{-d^*(q)(q-1)}  \;\;\;\, {\text{critical}}. \cr}
\end{equation}
Here $d^*(q)<d$ is the fractal dimension. Its dependence on $q$
is the hallmark of multifractality of wave functions. The
multifractal wave functions are delocalized, but extremely
inhomogeneous occupying only an infinitesimal fraction of the
sample volume in thermodynamic limit. The IPR is affected by
mesoscopic fluctuations which scale in metallic samples
as~\cite{mirlinprb} $\delta I_{\alpha}(2) \sim 1/g^2 \propto
L^{4-2d}$. In the critical region ($g \sim 1$)
fluctuations~\cite{mirlinprb} are of the same order as the
average value, which is then not enough to characterize the
critical eigenstates (even though multifractal wave
function extend throughout the whole sample, their IPR is not
self-averaging~\cite{evers}).
\begin{figure}
\centerline{ \psfig{file=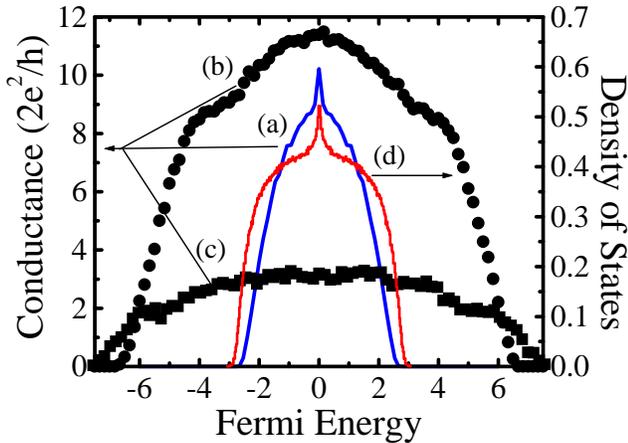,width=2.3in,angle=-90} }
\vspace{0.2in} \caption{Conductance and density of states in the
RH and DD Anderson models on a simple cubic lattice $16^3$: RH
disorder, (a) and (d), of strength $W_{\text{RH}}=1$ (mobility
edge is at $|E_c| \simeq 2.53$); diagonal disorder of strength
(b) $W_{\text{DD}}=6$ ($|E_c| \simeq 6.5$), and (c)
$W_{\text{DD}}=10$ ($|E_c| \simeq 7.4$). Disorder averaging is
performed over $N_{\text{Ens}}=20$ different samples for
conductance and $N_{\text{Ens}}=40$ for DoS.} \label{fig:dos}
\end{figure}

I use $I(2)$ as a rough guide in selecting eigenstates with
different properties in the delocalized phase (Fig.~\ref{fig:ipr}). 
The second parameter used in the selection procedure is the conductance
$g(E_F)$ computed for a band filled up to the Fermi energy $E_F$
equal to the state eigenenergy (see Fig.~\ref{fig:dos}). The
conductance as a function of band filling allows one to delineate
delocalized from localized phase as well as to narrow down the
critical region around LD transition point (which is defined by
$g(E_c) \sim 1$). Upon inspection of these two parameters, a
small window is placed around chosen energy, and $f(t)$ is
computed for all eigenenergies whose eigenvalues fall inside the
window. This provides more detailed information on the structure
of eigenstates than is encoded in IPR.
\begin{figure}
\centerline{ \psfig{file=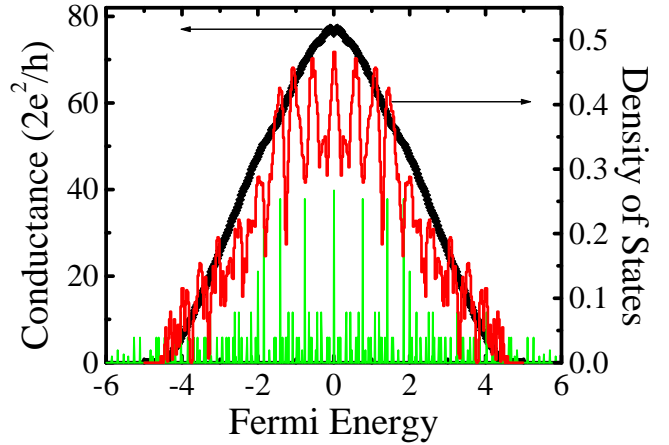,width=2.3in,angle=-90} }
\vspace{0.2in} \caption{Conductance and density of states in the
RH Anderson models on a simple cubic lattice $16^3$. The strenght of 
the RH disorder is $W_{\rm RH}=0.25$, which is so weak that features 
of a discrete spectrum of a finite-size lattice are still visible 
in the disorder-averaged DOS. Sharp lines correspond to the DoS of a
clean system on the same lattice (scaled by $1/10$ for clarity).
Disorder averaging is performed over $N_{\text{Ens}}=20$
different samples for conductance and $N_{\text{Ens}}=40$ for
DoS.} \label{fig:dos_wrh}
\end{figure}

The average IPR for both RH and DD Anderson model is shown in
Fig.~\ref{fig:ipr}. The models with random hopping  have been
around for a long time,~\cite{oppermann,inui} but have attracted
considerable attention only recently, inasmuch as they show a
disorder induced quantum critical point in less than three
dimensions,~\cite{viktor,brouwer,eilmes,fabrizio} where
delocalization occurs in the band center. Furthermore, several
models used for strongly correlated electron systems share
similar mathematical structure with the random hopping 2D models
formulated in the framework of (non-interacting) disorder
electron physics.~\cite{correlated} Real solids which could be
described by TBH~(\ref{eq:tbh}) with off-diagonal disorder
include doped semiconductors,~\cite{inui} such as P-doped Si,
where hopping integrals $t_{\bf mn}$ vary exponentially with the
distances between the orbitals they connect, while diagonal
on-site energies $\varepsilon_{\bf m}$ are nearly constant. The
behavior of low-dimensional RH Anderson model goes against the
standard mantra of the scaling theory of localization for
non-interacting systems~\cite{gang4} that all electron states are
localized in $d \leq 2$ for arbitrarily weak disorder. The
possibility of delocalization transition in one
dimension~\cite{cohen} goes back to the work of Dyson~\cite{dyson}
on glasses. Also, the scaling theory for quantum wires with
off-diagonal disorder requires two parameters~\cite{furusaki}
which depend on the microscopic model, thus breaking the
celebrated universality in disordered electron problems. In 3D
case explored here, the states in the band center are less
extended than other delocalized states inside the band
(Fig.~\ref{fig:ipr}). However, the off-diagonal disorder is not
strong enough~\cite{schreiber} to localize all states in the
band, in contrast to the usual case of diagonal disorder where
whole band becomes localized~\cite{slevin} for $W_{\text{DD}}^c
\gtrsim 16.5$.

The mobility edge for the strongest RH disorder
$W_{\text{RH}}=1$, as well as for DD models, is found by looking
at an exact zero-temperature static conductance. This quantity
(which is a Fermi surface property) is computed from the
Landauer-type formula~\cite{lb}
\begin{equation}\label{eq:landauer}
  g(E_F) =  \text{Tr} \, [{\bf t}(E_F)
  {\bf t}^{\dag}(E_F)],
\end{equation}
where transmission matrix ${\bf t}(E_F)$ is expressed in terms of
the real-space (lattice) Green functions~\cite{datta,nikolic} for
the sample attached to two clean semi-infinite leads. To study
the conductance in the whole band of the DD model, $t_{{\bf m}
{\bf n}}=1.5$ is used~\cite{nikolic} for the hopping integral in
the leads. This mesoscopic computational technique ``opens'' the
sample, thereby smearing the discrete levels of an initially
isolated system. Thus, the spectrum of {\it sample+leads=infinite
system} becomes continues and conductance can be calculated at
any $E_F$ inside the band. However, the computed conductance, for
not too small disorder~\cite{nikolic,mackinnon} or coupling to the
leads~\cite{hans} (which are of the same transverse width as the
sample~\cite{mackinnon}) is practically equal to the
``intrinsic'' conductance $g=\pi E_{\rm Th}/\Delta$
determined by the spectral properties of a closed sample.

The conductance and density of states (DoS)
\begin{equation}\label{eq:dos}
   N(E) =  2 \frac{\rho(E)}{V},
\end{equation}
are plotted in Fig.~\ref{fig:dos} for the samples whose
eigenstates are investigated (the factor of two is for spin
degeneracy). The DoS is obtained from the histogram of the number
of eigenvalues which fall into equally spaced energy bins along
the band. The conductance and DoS of the RH model have a peak at
$E=0$, which becomes a logarithmic singularity in the limit of
infinite system size.~\cite{dyson} To get an insight into the
``general weakness'' of the off-diagonal disorder,
Fig.~\ref{fig:dos_wrh} plots $g(E_F)$ and $N(E)$ for the low
$W_{\text{RH}}=0.25$. In this case $N(E)$ still resembles the DoS
of a clean system, even after ensemble averaging. On the other
hand, the conductance is a smooth function of energy since
discrete levels of an isolated sample are broadened by the
coupling to leads. The same is true for the DoS computed from the
imaginary part of the Green function for an open system. The
mobility edge is absent at low RH disorder ($W_{\rm RH}=0.25$ and
$W_{\rm RH}=0.375$) for system sizes $L \le 16a$. This means that
localization length $\xi$ is greater than $16a$ for all energies
inside the band of these systems. For other samples on
Fig.~\ref{fig:dos} the mobility edge appears inside the band.
This is clearly shown for $W_{\rm RH}=1$ case where band edge
$E_b$ ($N(E_b)=0$) differs from $E_c$. The mobility edge is
located at the minimum energy $|E_c|$ for which $g(E_c)$ is still
different from zero. The conductance of finite samples is always
finite, although exponentially small at $|E_F| > |E_c|$. The
approximate values of $|E_c|$ listed in Fig.~\ref{fig:dos} are
such that conductance satisfies: $g(E_F)<0.1$, for $|E_F|>|E_c|$;
typically $g(E_c) \in (0.2,0.5)$ is obtained, like in the recent
detailed studies~\cite{markos} of conductance properties at
$E_c$. Thus found $E_c$ is virtually equal to the true mobility
edge, which is properly defined only in thermodynamic limit (and
usually obtained from some numerical finite-size scaling
procedure~\cite{kramer}). Namely, the position of mobility edge
extracted in this way will not change~\cite{verges} when going to
larger system sizes if $\xi < L$ for all energies $|E| > |E_c|$.

The distribution $f(t)$ of  eigenfunction intensities has been
studied analytically for diffusive conductors close to the
universal RMT limit (where conductance is large and localization
effects are small) in Refs.~\onlinecite{mirlin,efetov} using the
supermatrix $\sigma$-model (NLSM),~\cite{efetovbook} or by means
of a ``direct optimal fluctuations method'' of
Ref.~\onlinecite{smolyarenko}. Numerical studies of statistics of
eigenstates in DD Anderson model were conducted in 2D for all
disorder strengths,~\cite{muller} while in 3D the focus has been
on the states appearing in the semiclassical transport regime
where comparison with analytical predictions (parameterized by
semiclassical quantities, like $k_F\ell$) can be made in the
regions of small~\cite{uski2p} and large~\cite{rare} deviations
of $f(t)$ from PT distribution. Here I show how $f(t)$ evolves
with the strength of (different types of) disorder in 3D samples,
where genuine LD transition occurs in the strong coupling regime
of the corresponding field-theoretical formulation.

In the weakly disordered ($k_F\ell \gg 1$) metallic ($g \gg 1$)
conductors pre-localized states are extremely rare. For example,
in an ensemble of 20 000 samples ($W_{\rm DD}=4$), whose typical
transport properties are well-described by semiclassical
theories, only four states would show up in the band center which
exhibit similar amplitude splashes like the one in
Fig.~\ref{fig:state}. Thus, to get a far tail (where deviations
from PT distribution are large) of $f(t)$ in such conductors one
has to search through enormous ensemble and locate special
configurations of a random potential.~\cite{rare} The maximum
wave function amplitudes which can be observed in this pursuit
are, plausibly, determined by the strength of disorder, i.e.,
conductance $g$. It is, however, interesting that tails at small
but finite $g$ in the strong disorder (like $W_{\rm DD}=10$) can
be longer than the tails of states in the localized phase, where
$g$ is vanishingly small, for $|E|>|E_c|$ at some smaller fixed
disorder (like in the case of $W_{\rm DD}=6$; compare the two
panels in Fig.~\ref{fig:ftdd} using respective conductances from
Fig.~\ref{fig:dos}). For strong enough disorder long tails of
$f(t)$ are found, even by investigating small ensembles of
disordered conductors (as shown below), since the frequency of
appearance of pre-localized states is greatly enhanced (while
system is still on the delocalized side of LD transition).

The complete eigenproblem of a single particle random Hamiltonian
is solved exactly by numerical diagonalization. Then, $f(t)$ is
computed as a histogram of intensities for the chosen eigenstates
in: delocalized ($|E|<|E_c|$), localized phase ($|E|>|E_c|$), and
critical region around the mobility edge $|E_c|$. The two delta
functions in Eq.~(\ref{eq:ft}) are approximated by a box function
$\bar{\delta}(x)$. The width of $\bar{\delta}(E-E_{\alpha})$ is
small enough at a specific energy that $\rho(E)$ is constant
inside that interval. For each sample, 5--10 states are picked by
the energy bin, which effectively provides additional averaging
over the disorder (according to ergodicity~\cite{ghur} in RMT).
The amplitudes of wave functions are sorted in the bins defined
by $\bar{\delta}(t-|\Psi_\alpha({\bf r})|^2 V)$ whose width is
constant on a logarithmic scale. The function $f(t)$ is computed
at all points inside the sample, i.e., $N=16^3$ in
Eq.~(\ref{eq:ft}).

The evolution of $f(t)$, when sweeping the band through the
interesting states, is plotted in
Figs.~\ref{fig:ftrh1},~\ref{fig:ftrh2} for the RH disordered
sample. Since pre-localized states generate slow decay of $f(t)$
at high wave function intensities (where PT distribution is
negligible),~\cite{efetovbook} this region is enlarged on
Fig.~\ref{fig:ftrh2}. This is obvious from the pre-localized
example in Fig.~\ref{fig:state} where state with large amplitude
spikes, highly unlikely in the framework of RMT, was found in a
very good metal. The same is trivially true for the localized
states which determine extremely long tails of $f_\xi(t)$. Thus,
the long asymptotic tails of $f(t)$, appreciably deviating from PT
distribution, are signaling the onset of localization. It is
interesting that states in the band center of RH model, which
define the largest zero-temperature conductance $[ g(E_F=0)
\approx 10.2$, Var $g(E_F=0)\approx 0.63 ]$, are in fact mostly
pre-localized. Moreover, both the frequency of their appearance
and high amplitude splashes resemble the situation at criticality
(NLSM-type calculation~\cite{fabrizio} shows that 3D wave
functions, sufficiently close to the band center, are always
extended for any disorder strength). It might be conjectured that
these pre-localized states would generate multifractal scaling of
IPR in the band center. This result, together with the DoS and
conductance from Figs.~\ref{fig:dos},~\ref{fig:dos_wrh}, shows
that phenomena in the band center of 3D conductors with
off-diagonal disorder are as intriguing as their much studied
counterparts in low-dimensional
systems.~\cite{viktor,brouwer,eilmes} The origin of these
phenomena can be traced back to a special sublattice, or
``chiral'',~\cite{inui,altland1} symmetry obeyed by
TBH~(\ref{eq:tbh}) with random hopping and constant on-site
energy (leading to an eigenspectrum which for $E_\alpha$ contains
$-E_\alpha$, with a special role played by $E=0$). In the
$W_{\text{RH}}=0.25$ and $W_{\text{RH}}=0.375$ cases all states
are extended. Their $f(t)$ looks similar to the distribution
function for the delocalized states at $E=1.5$ in the sample
characterized by $W_{\text{RH}}=1$. The distribution function
$f_{\xi}(t)$ defined in Eq.~(\ref{eq:ftexp}) fits reasonably well
the numerical $f(t)$ generated by the states around $E=2.75$,
where $\xi \simeq 1.2a$ is extracted for the localization length.
Thus, one can measure approximately $\xi$ in this
way~\cite{muller} even though the structure of localized
eigenstates can be more complicated~\cite{kramer} than the simple
radially symmetric exponential decay used to derive $f_{\xi}(t)$
[e.g., in the case of DD disorder $\xi \simeq 1.3a$ is obtained
for examined localized states in Fig.~\ref{fig:ftdd} for $W_{\rm
DD}=10$ ensemble, while corresponding states in $W_{\rm DD}=6$
ensemble seem to be too close to the mobility edge to follow
$f_\xi(t)$].

The same statistical analysis is performed for the eigenstates of
DD Anderson model---a ``standard model'' in the localization
theory. Figure~\ref{fig:ftdd} plots $f(t)$ at specific energies
$E_i$ in samples characterized by different conductances
$g(E_F=E_i)$. The conductance $g(E=0)$ of TBH with
$W_{\text{DD}}=6$ is numerically close to the conductance of RH
disordered samples with $W_{\text{RH}}=1$. Nevertheless,
comparison of the corresponding distribution functions reveals
disorder-dependent features~\cite{mirlin} which are beyond
universal corrections (i.e., independent of the details of random potential)
 accounted by the properties of a classical diffusion
operator~\cite{fmmodes,prigodin} (the spectrum of $-{\mathcal D}
\nabla^2$, with appropriate boundary conditions, depends on $g$,
shape of the sample and dimensionality; note that conductors at
$W \simeq 6$ with half-filled band lie on the boundary of
applicability of such semiclassical concepts~\cite{allen}). To get
the far tail of the eigenstate statistics in the band center a
much larger statistics is needed than used here.~\cite{rare} This
then makes the observed $f(t)$ in a special case of the band
center of $W_{\text{RH}}=1$ disordered conductor even more
spectacular because of very large amplitudes found in the small
ensemble of conductors. Thus, the strong dependence on the
microscopic details of random potential demonstrated in this
study is somewhat different from the disorder-specific short
length scale (``ballistic''~\cite{mirlin})
contributions~\cite{mirlin,rare} to the standard picture of
diffusive NLSM. Namely, here it seems that special features of
off-diagonal disorder in the band center generate completely
different functional form of the far tail, and not just some
disorder-specific values of the parameters~\cite{mirlin,rare} in
the exp-log-cube asymptotics.~\cite{mirlin,efetov,smolyarenko}.

In both models, all computed $f(t)$ intersect  PT distribution
(from below) around $6 \lesssim t \lesssim 10$, and then develop
tails far above PT values. The length of the tails is defined by
the largest amplitude exhibited in the pre-localized state (like
that in Fig.~\ref{fig:state}). For strong DD, $W_{\text{DD}}=10$,
the conductance $g(E_F)$ is smaller than $3.5$. In this regime
transport becomes ``intrinsically diffusive'',~\cite{allen} but
one can still extract resistivity from the approximate Ohmic
scaling of disorder-averaged resistance~\cite{allen} (for those
fillings where~\cite{todorov} $g(E_F)>2$). However, the close
proximity to the critical region $g \sim 1$ induces long tails of
$f(t)$ at all energies throughout the band---a sign of increased
frequency of appearance of highly inhomogeneous states. This
provides an insight into the microscopic structure of eigenstates
which carry the current in a non-semiclassical transport
regime~\cite{efetovbook} (characterized by the lack of simple
intuitive concepts, like mean free path $\ell$, since unwarranted
use of the Boltzmann theory would give~\cite{allen} $\ell < a$ in
this transport regime although the sample is still far away from
the LD transition).

In conclusion, the statistics of eigenstates in 3D mesoscopic
conductors, modeled by the tight-binding Hamiltonian on the
finite-size simple cubic lattice, have been studied. The disorder
is introduced either in the potential energy (diagonal) or in the
hopping integrals (off-diagonal). Also calculated are the average
inverse participation ratio of eigenfunctions and the exact
zero-temperature conductance as a function of Fermi energy. This
comprehensive set of parameters makes it possible to compare the
eigenstates in nanoscale samples with different types of
disorder, but characterized by similar values of conductance.
Disorder-specific details, which are not parameterized by the
conductance alone, are found. This is in spite of the fact that
dimensionality, shape of the sample, and conductance are expected
to determine the finite-size corrections to the universal
predictions of random matrix theory, at least in the samples which
are in the semiclassical transport regime. The appearance of
states with large amplitude spikes on the of top of RMT like
background is clearly demonstrated even in good metals. At
criticality, the proliferation of such ``pre-localized'' 
states is directly related to the extensively studied 
multifractal scaling of IPR. However, even in the delocalized 
phase with good metallic properties ($g \gg 1$), where the correlation 
length~\cite{jansen} $\xi_c \sim \lambda_F$ defined by the sample 
conductance $g(\xi_c) \sim 1$ is microscopic ($L<\xi_c$ would naturally 
account for the multifractal scaling,~\cite{jansen} like in 2D), 
pre-localized state with unexpectedly high amplitudes are 
found in the band center of random hopping disordered systems. 
They are inhomogeneous enough to generate extremely long tails 
of the distribution of eigenfunction amplitudes, akin to the ones
observed at criticality.

Inspiring discussions with V. Z. Cerovski are acknowledged.
Valuable guidance have been provided by P. B. Allen. The
important improvements of the initial cond-mat preprint have
resulted from criticism provided by B. Shapiro and I. E.
Smolyarenko. This work was supported in part by NSF grant no. DMR
9725037.

$\ddag$ Present address: Department of Physics, Georgetown
University, Washington, DC 20057-0995.


\begin{references}
\bibitem{crit} The usage of the term ``pre-localized'' (or anomalously localized)
in 3D systems is somewhat ambiguous, e.g., compare
Refs.~\onlinecite{jansen,mirlin,efetov,smolyarenko}.
\bibitem{foot} Note that random Hamiltonians of real disordered solids are tied to
the real-space representation (i.e., matrix elements in this
representation are spatially dependent, and e.g.,
TBH~(\ref{eq:tbh}) is a band diagonal matrix). Therefore, they do
not satisfy the statistical assumptions of standard RMT ensembles
since all elements of random matrices in RMT are non-zero and
spatially independent.~\cite{fyodorov} Nevertheless, the
connection to RMT statistics is provided by Efetov's
supersymmetric approach.~\cite{efetovbook}
\end{references}
\end{document}